\documentclass[12pt]{article}
\usepackage{amsmath} 
\usepackage{amssymb}
\usepackage{hhline}
\usepackage[scale={.75,.75}]{geometry}
\usepackage{url}
\usepackage{lscape}
\usepackage{caption}
\usepackage[numbers, sort&compress]{natbib}
\usepackage{epsfig}
\usepackage{multirow} 
\usepackage{color}
\usepackage{verbatim}
\usepackage{nicefrac}
\usepackage{upgreek}
\usepackage{bbm}
\usepackage{setspace}
\usepackage{pstricks}
\usepackage{psfrag}
\usepackage{color}

\usepackage{hyperref}

\newcommand{\p}{\textbf{p}}

\definecolor{green}{rgb}{0,0.5,0}

\begin{document}
\date{}

\title{\vspace{-2.5cm} 
\begin{flushright}
\vspace{-0.4cm}
{\scriptsize \tt TUM-HEP-949/14, NSF-KITP-14-059}  
\end{flushright}
{\bf On finite density effects}\\{\bf on cosmic reheating and moduli decay}\\{\bf and implications for Dark Matter production
}
}

\author{Marco Drewes\\ 
\footnotesize{Physik Department T70, Technische Universit\"at M\"unchen, }\\
\footnotesize{James Franck Stra\ss e 1, D-85748 Garching, Germany}}

\maketitle

\vspace{-0.5cm}
\begin{abstract}
  \noindent 
We study the damping of an oscillating scalar field in a Friedmann-Robertson-Walker spacetime by perturbative processes, taking into account the finite density effects of interactions with the plasma of decay products on the damping rate. The scalar field may be identified with the inflaton, in which case this process resembles the reheating of the universe after inflation. It can also model a modulus that dominates the energy density of the universe at later times. We find that the finite density corrections to the damping rate can have a drastic effect on the thermal history and considerably increase both, the maximal temperature in the early universe and the reheating temperature at the onset of the radiation dominated era. As a result the abundance of some Dark Matter candidates may be considerably larger than previously estimated. We give improved analytic estimates for the maximal and the reheating temperatures and confirm them numerically in a simple model.
\end{abstract}

\section{Introduction}
Many properties of the cosmos we observe today are the result of processes that occurred during the early phase of its history, during which it was filled with a hot primordial plasma \cite{Kolb:1990vq}. 
This makes the thermal history of the universe crucial for physical cosmology. 
The Cosmic Microwave Background (CMB) radiation, which was emitted when the temperature was about $T\sim0.25$ eV, provides the earliest direct probe of the high temperature phase.
Its temperature fluctuations indirectly carry information about earlier times. 
The light elements in the intergalactic medium also provide an probe of earlier times. They were created in thermonuclear reactions in the primordial plasma, known as big bang nucleosynthesis (BBN). The good agreement between theoretical BBN calculations and observation allows to conclude that the standard picture of cosmology holds at least up to $T\sim 1$ MeV. Observationally, very little is known about the time before that.
However, there are good reasons to believe that the universe has been exposed to much higher temperatures. 
There is overwhelming evidence that most of the mass in the observable universe is composed of non-baryonic Dark Matter (DM). If the DM particles were produced thermally, then the temperature should have been at least comparable to their mass. 
If the observed baryon asymmetry of the universe\footnote{See e.g. \cite{Canetti:2012zc} and references therein for a detailed discussion.} was caused by baryon number (B) violating thermal electroweak sphalerons \cite{Kuzmin:1985mm}, a temperature $T_{EW}\gtrsim 140$ GeV is required \cite{Burnier:2005hp,D'Onofrio:2014kta}.\footnote{See \cite{GarciaBellido:1999sv,Brauner:2011vb,Tranberg:2012jp} for suggestions to circumvent this bound.}  
The overall geometry of the universe and the properties of CMB temperature fluctuations suggest that it underwent a period of \emph{cosmic inflation}, i.e. accelerated cosmic expansion, at very early times \cite{Starobinsky:1980te,Guth:1980zm,Linde:1981mu}. Grand unified theories tend to relate the scale of inflation to an energy scale $\sim 10^{16}$ GeV (while the inverse conclusion that a high scale of inflation implies grand unification is of course not true \cite{Lyth:2014yya}). If confirmed, the observations of the BICEP2 telescope \cite{Ade:2014xna} support this idea.

If inflation is driven by the potential energy of a scalar inflaton field $\phi$, then the energy in other degrees of freedom gets diluted away, leaving a cold and empty universe in which all energy is contained in the zero mode of the "classical" (mean) field $\langle\phi\rangle$. It is released into all other degrees of freedom at a rate $\Gamma$ by dissipative effects during the oscillations of $\langle\phi\rangle$ around the minimum of its potential $V(\phi)$ \cite{Traschen:1990sw,Shtanov:1994ce,Kofman:1994rk,Kofman:1997yn,Boyanovsky:1994me}. 
This process initiates the radiation dominated era of cosmic history, which is described by standard big bang cosmology. The initial temperature of this era, the \emph{reheating temperature}, is set by the details of the mechanism by which $\phi$ dissipates its energy. It is, however, in general not the largest temperature the universe has ever been exposed to, which is usually reached before the universe becomes radiation dominated \cite{Giudice:2000ex}. 

Alternatively the energy density of the universe after inflation may be dominated by other scalar fields. Such \emph{moduli} appear in many realisations of string theory. In this case the radiation dominated era would commence once the moduli decay. 
In many extensions of the Standard Model (SM) of particle physics the lifetime of the predicted moduli is so long that its late decay would be in conflict with observations, leading to the \emph{moduli problem} \cite{Coughlan:1983ci,Goncharov:1984qm,German:1986ds,Ellis:1986zt,Banks:1993en,Randall:1994fr}, but it can also lead to interesting alternative cosmic histories, including non-thermal DM production \cite{Watson:2009hw,Acharya:2008bk,Acharya:2009zt}.
In both, the standard reheating scenario and histories involving moduli, the thermal and history of the universe are strongly affected by the dynamics and decay of one or several scalar fields.

A quantitative understanding of this process is crucial for physical cosmology, as the time evolution of the temperature strongly affects the abundance of thermal relics. If all degrees of freedom were in thermal equilibrium after inflation, then the only relevant quantity is the maximal temperature in the early universe
Most models of baryogenesis rely on temperatures much above $T_{EW}$. 
Standard leptogenesis \cite{Fukugita:1986hr}, for instance, requires $T> 10^{8}$ GeV \cite{Davidson:2002qv}, which leads to the gravitino overproduction problem in supersymmetric theories \cite{Pagels:1981ke}. 
This tension can be eliminated if the scale of Majorana masses is below the electroweak scale \cite{Akhmedov:1998qx,Drewes:2012ma}, in which case leptogenesis can be successful for $T< 10^7$ GeV\footnote{Further reduction is possible if the right handed neutrinos' Majorana masses are degenerate \cite{Pilaftsis:2003gt,Asaka:2005pn,Shaposhnikov:2008pf,Racker:2012vw,Canetti:2012vf,Canetti:2012kh,Shuve:2014zua}.} and may be probed in the laboratory \cite{Canetti:2012vf,Canetti:2014dka}.
If some constituents of the plasma do not reach thermal equilibrium before freezeout, 
then their abundance is not only sensitive to the maximum temperature, but to the details of the thermal history \cite{Giudice:2000ex,Erickcek:2011us,Roszkowski:2014lga,Harigaya:2014waa}.

It is common to estimate the reheating temperature by assuming that $\phi$ dissipates all its energy into radiation instantaneously when its vacuum decay rate $\Gamma_0$ equals the rate $H$ of Hubble expansion.
However, in reality the heating process is much more complicated.
In many (but not all) models it is initiated by a preheating phase, in which non-perturbative particle production is strongly enhanced by a parametric resonance \cite{Traschen:1990sw,Kofman:1994rk,Kofman:1997yn}.
Regardless of the existence of the resonance $\phi$ always dissipates energy by perturbative processes, which becomes the main heating mechanism at late time even if initially non-perturbative particle production dominated.

In this work we are only concerned with perturbative reheating, which is simpler to describe and sets the temperature at the onset of the radiation dominated era.
We show that even if reheating is entirely driven by perturbative processes, the use of $\Gamma_0$ can lead to a significant underestimate of the reheating temperature and maximal temperature.
The reason is that the dissipation rate $\Gamma$ at which $\phi$ transfers energy into radiation at a given point in time is affected by the radiation that has been produced at earlier times.
This radiation forms a hot plasma, and the interaction with this plasma can strongly modify $\Gamma$.
This has already been pointed out in \cite{Kolb:2003ke}.
However, the authors restricted the analysis to the effect of "thermal masses" on $1\rightarrow 2$ decays and neglected the (usually more important) quantum statistical effects and scatterings.
This shortcoming was pointed out in \cite{Yokoyama:2005dv,Yokoyama:2006wt,Bodeker:2006ij} and studied in more detail in \cite{Mukaida:2012qn,Mukaida:2012bz,Drewes:2013iaa}.\footnote{Similar thermal effects have also been discussed in the context of the curvaton scenario \cite{Enqvist:2012tc,Enqvist:2013qba,Enqvist:2013gwf,Kitajima:2014xna} and are crucial for the idea of warm inflation \cite{BasteroGil:2010pb,BasteroGil:2012cm}.} 
In this work we use the results of \cite{Drewes:2013iaa} to study the time evolution of the temperature during perturbative reheating. We find that the enhancement of the full dissipation rate $\Gamma$ due to induced transitions and scatterings at finite density can lead to considerably higher temperatures in the early universe than the commonly used estimates based on the vacuum decay rate $\Gamma_0$ would suggest.
As a result, the abundance of thermally produced relics  including DM can be considerably larger than previously estimated.
Our analysis can be applied to both, the inflaton or a moduli field that dominates the energy density, therefore we will in the following not specify the role of $\phi$.

\section{General analysis}
In the following we consider a generic scalar field $\phi$ with a large mass $m_\phi$ that performs coherent oscillations near the minimum of its potential $V(\phi)$ in a Friedmann-Robertson-Walter spacetime with scale factor $a$ and Hubble parameter $H\equiv\frac{d a}{d t}/a$. 
Initially all energy is stored in the zero mode of the oscillating mean field (or one-point function) $\langle\phi\rangle$.
Its interactions with other degrees of freedom, the strength of which we shall characterise by a dimensionless number $\alpha$, lead to dissipation that damps the oscillation and transfers energy to these degrees of freedom.
This is a far from equilibrium process in a time-dependent background and possibly dense medium formed by the decay products. In such a situation the standard methods to calculate S-matrix elements in particle physics cannot always be applied because there is no well-defined notion of asymptotic states, the properties of propagating states may significantly differ from those of particles in vacuum and classical particle number in general is not a suitable quantity to characterise the system. A suitable framework is offered by the Closed-Time-Path (CTP) formalism of nonequilibrium quantum field theory \cite{Schwinger:1960qe,Bakshi:1962dv,Bakshi:1963bn,Keldysh:1964ud}, in which all observables can be expressed in terms of time dependent correlation functions of the quantum fields without reference to asymptotic states or free particles. 
The equations of motion for the correlation functions are in general complicated second order integro-differential equations of the Kadanoff-Baym type \cite{KBE}.
However, under the assumptions made above the weakness of the coupling $\alpha$ implies a separation of time scales that justifies to treat the time evolution of the background as slow compared to the time scale $1/m_\phi$ \cite{Anisimov:2008dz,Drewes:2012qw}.
Then the dynamics can effectively be captured by (quantum) kinetic equations  of the Boltzmann type, i.e. differential equations for generalised distribution functions that are first order and local in space and time \cite{Drewes:2012qw}.\footnote{
This holds even at the level of loop corrections and/or when coherent (e.g. flavour) oscillations are important 
\cite{Cirigliano:2009yt,Anisimov:2010dk,Anisimov:2010aq,Beneke:2010dz,Fidler:2011yq,Frossard:2012pc,Dev:2014laa}. In the latter case the quantum kinetic equation is matrix-valued.
} The dissipation rates appearing in these equations, however, should be calculated in the CTP formalism.

\subsection{Boltzmann equations}
We assume that the $\phi$-mass $m_\phi$ is much larger than the masses of all known particles. 
This implies that the decay products are relativistic when produced, and we can effectively treat them as a bath of radiation with $g_*$ number of degrees of freedom. 
We will in the following assume that $g_*\gg1$ is constant at all times for simplicity to obtain analytic solutions. In reality this may be a bad approximation, as new particle species are continuously produced. However, the precise time evolution of $g_*$ is strongly model dependent and requires to take into account all production thresholds and the details of the thermalisation process.
The energy density $\rho_\phi$ of the (averaged) coherent $\langle\phi\rangle$-oscillations redshifts as $\propto a^{-3}$, i.e. like non-relativistic matter. 
The energy density $\rho_R$ of the radiation, on the other hand, redshifts as $\propto a^{-4}$.
In all what follows we parametrise $\rho_R$ and the time dependence of the dissipation rate $\Gamma$ by an effective temperature $T$. 
This is probably not a good approximation if non-perturbative particle production and parametric resonance are at work during an early preheating phase, in which the plasma has no time to thermalise. 
The thermalisation of the plasma is a highly non-trivial process, 
which in the context of inflation has e.g. been addressed in Refs.~\cite{Ellis:1987rw,Dodelson:1987ny,Enqvist:1990dp,Enqvist:1993fm,McDonald:1999hd,Davidson:2000er,Berges:2010zv,Mazumdar:2013gya,Harigaya:2013vwa,Harigaya:2014waa} and references therein. 
However, our approximation should capture the main effects if the universe is predominantly heated by perturbative processes. Then the plasma has time to at least partially reach kinetic equilibrium, as the constituents of the bath usually have much stronger interactions amongst each other than with $\phi$.\footnote{For a moduli this is the case because the decay happens at late times. Also for the inflaton the interactions are usually assumed to be be very feeble, as its effective potential has to be sufficiently flat to maintain slow roll inflation.}
We express the strength of a typical interaction amongst the bath constituents by a dimensionless number $1>\lambda\gg \alpha$, which could e.g. be a gauge coupling constant.
The use of an effective temperature is a good approximation for the late decay of a moduli in non-thermal histories that occurs long after SM gauge interactions came into equilibrium.

Under these assumptions one can perform an averaging over momenta. 
Hence, a simple set of momentum averaged Boltzmann equations is sufficient for our purpose.
\begin{eqnarray}
\frac{d \rho_\phi}{d t}+3H\rho_\phi+\Gamma\rho_\phi&=&0\label{BE1}\\
\frac{d \rho_R}{d t}+4H\rho_R-\Gamma\rho_\phi&=&0\label{BE2}
\end{eqnarray}
Following the above "derivation" it is, however, clear that $\Gamma$ should be calculated from first principles in the CTP-formalism or thermal field theory.
 For simplicity we ignore the fact that the dissipation will also populate other modes of $\langle\phi\rangle$ 
 or produce (possibly non-relativistic) $\phi$-particles, which in the CTP-formalism would be captured by the two-point (and possibly higher order) correlation functions of $\phi$. This is physically well justified by the large number of degrees of freedom $g_*$ in the bath.
It is convenient to introduce the variables $\Phi\equiv\rho_\phi a^3/m_\phi$, $R\equiv\rho_R a^4$ and $x\equiv a m_\phi$, in terms of which we can rewrite (\ref{BE1}) and (\ref{BE2}) as
\begin{eqnarray}
\frac{d\Phi}{dx}&=&-\frac{\Gamma}{Hx}\Phi\label{be1}\\
\frac{dR}{dx}&=&\frac{\Gamma}{H}\Phi\label{be2}
\end{eqnarray}
with
\begin{eqnarray}
H&=&\left(\frac{8\pi}{3}\right)^{1/2}\frac{m_\phi^2}{m_P}\left(\frac{R}{x^4}+\frac{\Phi}{x^3}\right)^{1/2}\label{Hubble}\\
T&=&\frac{m_\phi}{x}\left(\frac{30}{\pi^2g_*}R\right)^{1/4}\label{TofR},
\end{eqnarray}
were $m_P$ is the Planck mass. 

\subsection{Maximal temperature and reheating temperature}
Once $\phi$ has dissipated part of its energy into other degrees of freedom, it oscillates in a plasma formed by the decay products. 
This has several effects on the dissipation rate $\Gamma$. On one hand quantum statistics, i.e. Bose enhancement and Pauli blocking, modify the transition amplitudes into different final states. On the other hand the possibility of inelastic scatterings with quanta from the bath opens up new channels of dissipation. Finally, the properties of quasiparticles in the plasma can differ considerably form those of particles in vacuum. In particular, "thermal masses" make the phase space temperature dependent.
With the assumptions outlined above, these effects can be parametrised by treating $\Gamma$ as a function of $T$ (and thereby $R$).\footnote{In the most general set-up it also depends on $\langle\phi\rangle$ and hence $\Phi$. We neglect this dependence, assuming that the amplitude of the $\phi$-oscillations is not too large.}
A general temperature dependent damping rate can be Taylor expanded in powers of $T/m_\phi$.
\footnote{Here we assume that all other vacuum masses are much smaller than $m_\phi$, otherwise we would have to include powers of the ratios of all dimensionful parameters.}
Expanding around $T=0$ we obtain
\begin{equation} 
\Gamma=\sum_{n=0}^\infty \Gamma_n \left(\frac{T}{m_\phi}\right)^n \label{GammaExansion}
\end{equation}
In the following we study the time evolution of the temperature $T$ with the initial conditions $a=a_I=1/m_\phi$, $R=T=0$ and $\Phi=\Phi_I=V_I/m_\phi^4$, where $V_I$ is $V(\phi)$ at initial time. 
We define the \emph{maximal temperature} as the maximum of $T$  as a function of $x$ and the \emph{reheating temperature} as the temperature in the moment $\rho_\phi=\rho_R$. The latter corresponds to 
\begin{equation}
\frac{R}{x^4}=\frac{\Phi}{x^3}.\label{PhiRadEq}
\end{equation} 
and usually coincides in good approximation with the moment when $\Gamma=H$, as $\rho_\phi$ is dissipated within one Hubble time once this point is reached.

\subsubsection{Instantaneous reheating}
In a zeroth order approximation we can neglect the the time dependence of $\Gamma$, i.e. replace it by the vacuum decay rate $\Gamma_0$, 
 set the right hand sides of (\ref{be1}) and (\ref{be2}) to zero for $\Gamma<H$ and assume that $\phi$ instantaneously dumps all its energy into radiation when $\Gamma=H$.
Using (\ref{Hubble}) and (\ref{TofR}) with $\Phi=0$ we obtain the commonly used estimate
\begin{equation}
T_R\equiv\sqrt{\Gamma_0 m_P }\left(\frac{90}{8\pi^3g_*}\right)^{1/4}\label{TR}
\end{equation}
In this approximation the maximal and the reheating temperature are both identical to $T_R$.
In reality, the dissipation happens at a finite rate $\Gamma$. 
One could argue that $T_R$ provides an upper bound on the temperature in the early universe:
As the radiation cools down due to the universe's expansion during the time span $\sim 1/\Gamma$ that this process takes, instantaneous conversion of the energy should be the most efficient way heating. 
However, this argument is incorrect because dissipation already starts before the moment
$\Gamma=H$. Even though the fraction by which $\rho_\phi$ is reduced per
Hubble time is very small prior to $\Gamma=H$, the absolute amount of
energy released into radiation is larger than at $\Gamma=H$ and later
times because of the larger value of $\Phi$. 
As it has previously been observed in \cite{Giudice:2000ex}, the maximal temperature is usually considerably larger than $T_R$ and reached well before $\Gamma=H$.
This has the linguistically curious consequence that $T$ actually decreases during most of the reheating process.
The period between the maximal temperature and $\Gamma=H$ is, however, different from a ordinary matter dominated phase because $\rho_R$ decreases slower than $\propto a^{-4}$ due to the gain term on the right hand side of (\ref{BE2}), which partly compensates the cooling by Hubble expansion.

\subsubsection{Maximal temperature and reheating temperature without finite density corrections}\label{NoMediumCorr}
In a more realistic approach we can approximately solve (\ref{be1}) and (\ref{be2}) analytically. 
Before $\Gamma=H$ we can in good approximation set $\Phi=\Phi_I$ as constant and treat it as an external source in (\ref{BE2}). 
For $\Gamma=\Gamma_0$ it is easy to obtain the solution 
\begin{equation}
R=A_0\frac{2}{5}(x^{5/2}-1)\label{R0}
,\end{equation}
where $A_0$ is given by
\begin{equation}
A_n\equiv\frac{\Gamma_n}{m_\phi}\sqrt{\Phi_I}\frac{m_P}{m_\phi}\left(\frac{30}{\pi^2g_*}\right)^{n/4}\left(\frac{3}{8\pi}\label{An}\right)^{1/2}
,\end{equation}
which we defined for arbitrary $n$ for later use.
From this we obtain by using (\ref{TofR})
\begin{equation}
T=m_\phi \left(A_0\frac{30}{\pi^2g_*}\frac{2}{5}\right)^{1/4}\left(x^{-3/2}-x^{-4}\right)^{1/4}\label{T1}
.\end{equation}
The maximum of (\ref{T1}) is at $x_{max}=(8/3)^{2/5}$, hence we define
\begin{eqnarray}
T_{max}&\equiv& m_\phi \left(A_0\frac{30}{\pi^2g_*}\frac{2}{5}\right)^{1/4}\left(x_{max}^{-3/2}-x_{max}^{-4}\right)^{1/4}\\
&\simeq&0.6 \left( \frac{\Gamma_0}{g_*} m_P \right)^{1/4}V_I^{1/8} \simeq 0.7 T_R^{1/2}\left(\frac{V_I}{g_*}\right)^{1/8}\label{Tmax}
.\end{eqnarray}
After reaching $T_{max}$ the temperature decreases as $T\propto a^{-3/8}$ (with $x=a/a_I$) until reaching $T_R$ at $\Gamma=H$.
The difference to the usual relation $T\propto 1/a$ is due to the dissipation: For $x>x_{max}$ it is not sufficient to heat the universe and $T$ decreases with time, but the decrease is slowed down due to the dissipation term on the right hand side of (\ref{be2}). 
The energy density during this period is dominated by $\rho_\phi$, 
which in good approximation redshifts like non-interacting matter as long as $\Gamma\ll H$.
This implies $a\propto t^{2/3}$, $H\propto T^4$ and $T\propto t^{-1/4}$ (assuming constant $g_*$).
For $\Gamma>H$ the usual relations in the radiation dominated era hold, i.e. $T\propto 1/a$  with $a\propto t^{1/2}$ and $H\propto T^2$.
Figure \ref{NoThermal} shows that (\ref{Tmax}) indeed provides an excellent approximation to the maximal temperature if $\Gamma$ is time-independent.
At the same time the standard definition (\ref{TR}) under this assumption still gives a good estimate for the reheating temperature at the onset of the radiation dominated era. This is easy to understand, as it can be obtained from the expression for the radiation density in equilibrium, which is insensitive to the previous history.

\subsubsection{Construction of a general solution}\label{Sec:GeneralSol}
To this end our results agree with those found in Ref.~\cite{Giudice:2000ex}. 
They were obtained under the assumption of a time independent $\Gamma$, which can be expressed in terms of the criterion
\begin{equation}
\Gamma_0\gg \sum_{n=1}^\infty\Gamma_n\left(\frac{T}{m_\phi}\right)^n.\label{gamma0criterion}
\end{equation}
If the series in the relevant temperature regime is dominated by one term with coefficient $\Gamma_m\lesssim \Gamma_0$, i.e. $\Gamma(T)$ in good approximation is the sum of a constant term and a power law, (\ref{gamma0criterion}) can be rewritten as $T<(\Gamma_0/\Gamma_m)^{1/m}m_\phi$. This tends to hold if the main contributions to $\Gamma$ come from one interaction and finite density corrections to the quasiparticle dispersion relations ("thermal masses") are negligible.
These modify the phase space and lead to significant deviations from a power law, e.g. by introducing kinematic thresholds \cite{Kolb:2003ke,Yokoyama:2005dv,Drewes:2010pf,Drewes:2013iaa}.
The fact that thermal masses are usually not relevant for $T<m_\phi/\lambda$ is often used to argue that thermal effects are entirely negligible in this regime. However, quantum statistical effects due to Bose enhancement and Pauli blocking are already at work for $T\sim m_\phi$. Moreover, since the particles in the plasma are light, they can reach sizable occupation numbers even at $T<m_\phi$. This opens the possibility for new channels of dissipation due to interactions with quanta from the bath, such as Landau damping, which can give a dominant temperature dependent contribution to $\Gamma$ even at $T<m_\phi$ \cite{Drewes:2013iaa}.

As long as $\Gamma$ as a function of $T$ can piece-wise be approximated by a power law 
we can nevertheless find an approximate analytic solution for $R$ that allows to estimate the maximal temperature and reheating temperature.
We first solve (\ref{be2}) for $\Gamma=\Gamma_n T^n m_\phi^{-n}$ with constant $\Phi=\Phi_I$ and initial condition $R=R_i$ at $x=x_i$, which is defined as the moment when $T=T_i$. The solution is
\begin{equation}
R=\left(A_n\frac{1-n/4}{5/2-n}\left( x^{5/2-n} - x_i^{5/2-n} \right) + R_i^{1-n/4}\right)^{1/(1-n/4)}\label{Rn}
\end{equation}
Now we can construct a complete solution by matching. 
Let us assume that there is a finite number of temperature intervals, separated by temperatures $T_i$, in each of which the function $\Gamma(T)$ can be approximated by a power law. 
If we set the initial values $T=0$ and $R=R_0=0$ at $x=x_0=1$ we can use (\ref{R0}) for the period $x_0<x<x_1$. That solution evaluated at $x_1$ provides the boundary condition $R_1$ at $x_1$, which can be inserted into (\ref{Rn}) to find the solution for $x_1<x<x_2$. As long as $\Gamma(T)$ can piece-wise be approximated by a power law, we can iteratively construct an analytic solutions for all times.
The same procedure can of course be applied to non-zero initial $\rho_R$ by setting $R_0\neq0$ at $x_0$. This is in general the case if one is concerned with the dynamics of a moduli at late times or if significant amounts of radiation have been created in a preheating phase.
It may require the use of (\ref{Rn}) with $n>0$ already at $x_0$ if the initial temperature is high enough that (\ref{gamma0criterion}) does not hold.

\subsubsection{Maximal temperature and reheating temperature with finite density corrections}
We can explicitly perform this procedure for the case 
\begin{equation}
\Gamma=\Gamma_0+\Gamma_2 \frac{T^2}{m_\phi^2}.\label{GammaApprox}
\end{equation}
Here $\Gamma_0$ is the vacuum decay rate.
A term $\propto T^2$ appears quite generically in the high temperature regime, as one can see by simple dimensional analysis. The relaxation rate $\Gamma(\p,T)$ for a mode $p$ of $\phi$ is related to the imaginary part of the retarded $\phi$-self energy as 
\begin{equation}
\Gamma(\p,T)=-\frac{\mathcal{Z}_\p}{p_0}{\rm Im}\Pi^R(p,T)\Big|_{p_0=\Omega_\p} .\label{PiR}
\end{equation} 
Here $\Omega_\p$ is a quasiparticle mass shell, defined as solution of $p^2-m_\phi^2-{\rm Re}\Pi^R(p,T)=0$, and the residue is 
\begin{equation}
\mathcal{Z}_\p=\left[1-\frac{1}{2\Omega_\p}\frac{\partial {\rm Re}\Pi^R_\p(p_0)}{\partial p_0}\right]^{-1}_{p_0=\Omega_\p}.
\end{equation}
We are interested in the zero mode $\p=0$. 
Since we assume that $\phi$ has only feeble interactions, finite density corrections to its dispersion relation can be neglected and we simply use the vacuum mass $p_0=m_\phi$ and set the residue $\mathcal{Z}_\p=1$. In the limit $T\gg m_\phi$ dimensional analysis suggests that the leading contribution to $\Pi^R$ is $\propto T^2$, hence leading order contribution to (\ref{PiR}) is $\propto T^2/m_\phi$.
Indeed (\ref{GammaApprox}) provides a good approximation to the example given in section \ref{Sec:Model} in the regime where thermal masses are not too large. 
Using (\ref{gamma0criterion}), we define the temperature were finite density effects become important as $T_1\equiv m_\phi(\Gamma_0/\Gamma_2)^{1/2}$.  This allows to distinguish three different cases.

\paragraph{$T_R<T_{max}<T_1$:} If the temperatures $T_R$ and $T_{max}$ defined in (\ref{TR}) and (\ref{Tmax}) are both smaller than $T_1$, then they give good approximations for the reheating temperature and maximal temperature, see figure \ref{NoThermal}. 

\paragraph{$T_R<T_1<T_{max}$:} If $T_1<T_{max}$, then (\ref{Tmax}) cannot be used to estimate the maximal temperature the universe is exposed to.
In this case the temperature surpasses $T_1$ before reaching its maximum. We can use the matching procedure described above to obtain
\begin{eqnarray}
R&=&\theta(x_1-x) A_0\frac{2}{5}(x^{5/2}-1)
+ \theta(x-x_1)\left(A_2\left( x^{1/2} - x_1^{1/2} \right) + R_1^{1/2}\right)^{2}\label{R2}
.\end{eqnarray}
The solution (\ref{R2}) remains valid for $x>x_{max}$ and roughly implies $T\propto a^{-3/4}$ until the point $x_2$ when $T$ drops below $T_1$ again. 
The solution for later times is simply obtained by matching: $R$ at $x>x_2$ is given by (\ref{Rn}) with $n=0$ and boundary condition $R_2$ set by evaluating (\ref{R2}) at $x=x_2$.
For $x>x_2$ the temperature roughly scales as $T\propto a^{-3/8}$ until reaching $T_R$ at $\Gamma=H$ and subsequently enters the standard radiation dominated era, cf. section \ref{NoMediumCorr}.
In the applications we have in mind the initial amplitude of the $\phi$-oscillations is usually very large, hence $\sqrt{V_I}\gg m_\phi$ and $A_n\gg 1$, which allows to approximate
\begin{eqnarray}
x_1&\simeq&1 +\frac{2}{5} \left(\frac{\Gamma_0}{\Gamma_2}\right)^{1/2}\frac{5\pi^2g_*}{60A_0} \ , \ R_1=A_0\frac{2}{5}(x_1^{5/2}-1) 
.\end{eqnarray}
From this we can find that $T$ is maximal at $\tilde{x}_{max}\simeq (4/3)^2x_1$, when $R=\tilde{R}_{max}\simeq (A_2 /3)^2 x_1$. The maximal temperature is
\begin{eqnarray}
\tilde{T}_{max}&=&m_\phi x_1^{-3/4}\left(\frac{3}{4}\right)^2\left(\frac{A_2}{3}\right)^{1/2}\left(\frac{30}{\pi^2g_*}\right)^{1/4}\nonumber\\
&\simeq&0.33\sqrt{ m_P\frac{\Gamma_2}{g_*}}\frac{V_I^{1/4}}{m_\phi}\label{TtildeMax}
\end{eqnarray} 
Interestingly (\ref{TtildeMax}) to leading order in the small ratios $m_\phi/m_P$, $m_\phi/V_I$ etc. does not depend on $\Gamma_0$ in spite of the fact that the heating up to $T_1$ is governed by this parameter. This holds as long as $\tilde{T}_{max}$ lies considerably above $T_1$, then $x_1$ is very close to $1$ and most of the heating occurs while $T>T_1$. If the temperature drops below $T_1$  before $\Gamma=H$, then $T_R$ defined in (\ref{TR}) still gives a reliable estimate of the temperature at the onset of the radiation dominated era in spite of the fact that the maximal temperature $\tilde{T}_{max}$ is much larger than the "naive" estimate $T_{max}$.  
This is illustrated in figure \ref{TmaxFig}.

\paragraph{$T_1<T_R<T_{max}$:}
If $T_R$ and $T_{max}$ are both larger than $T_1$, then the radiation dominated ($\rho_\phi<\rho_R$) era starts in the regime where the temperature dependence of $\Gamma$ cannot be neglected. 
In this case one would expect that the maximal temperature is still given by (\ref{TtildeMax}) and the reheating temperature can be estimated by using the temperature dependent expression (\ref{GammaApprox}) when solving $\Gamma=H$, which gives the solution
\begin{equation}
\tilde{T}_R\equiv\left(
\frac{m_\phi \Gamma_0}{\left(\frac{8\pi^3g_*}{90}\right)^{1/2}\frac{m_\phi}{m_P}-\frac{\Gamma_2}{m_\phi}}
\right)^{1/2}.\label{TRtilde}
\end{equation}
Interestingly, the expression (\ref{TRtilde}) becomes imaginary for $T_R>T_1$,
which requires  $\Gamma_2>\Gamma_2^{\rm crit}$ with 
\begin{equation}
\Gamma_2^{\rm crit}= \frac{m_\phi^2}{m_P}\left(\frac{8\pi^3g_*}{90}\right)^{1/2}\label{Gamma2crit}
\end{equation}
This can easily be understood by comparing (\ref{Hubble}) with (\ref{TofR}) to (\ref{GammaApprox}): In a radiation dominated universe ($R/x\gg\Phi$) at $T>T_1$ the Hubble rate and $\Gamma$ both scale as $\propto T^2$, then $\Gamma_2>\Gamma_2^{\rm crit}$ implies $\Gamma>H$ at all times. 
At first this seems to be of little relevance, as our initial condition $\rho_r=0$ implies that the dissipation rate at initial time is of course given by $\Gamma=\Gamma_0$, which for any reasonable choice of parameters is smaller than the initial $H$. Hence, the moment $\Gamma=H$ is always reached at finite time.

However, it turns out that $\rho_\phi=\rho_R$ is reached much before $\Gamma=H$, i.e. at much higher temperatures.
To see this we consider the approximate solution (\ref{R2}). 
Usually (\ref{R2}) cannot be used to determine the beginning $\rho_R\sim\rho_\phi$ of the radiation dominated era: (\ref{PhiRadEq}) has no solution for $x>1$ if $R$ is given by (\ref{R2}). 
This is because (\ref{R2}) was obtained by keeping $\Phi=\Phi_I$ constant; in reality $\Phi$ of course slowly decreases with $x$ due to dissipation, an at some point the validity of (\ref{R2}) breaks down. 
However, for $\Gamma_2>\Gamma_2^{\rm crit}$ the approximate expression (\ref{R2}) grows sufficiently fast to catch up with $\Phi$ and solves (\ref{PhiRadEq}) at finite
\begin{equation}
x_{\rm crit}= \left(\frac{A_2\sqrt{x_1}-\sqrt{R_1}}{A_2-\Phi_I}\right)^2\label{xcrit}
.\end{equation}
 Reinsertion into (\ref{Hubble}) and (\ref{GammaApprox}) shows that $\Gamma\ll H$ at $x=x_{\rm crit}$. This suggests that the universe becomes radiation dominated long before $\Gamma=H$, and (\ref{TRtilde}) is not at all a good estimate for the reheating temperature. 
Our numerical solution of (\ref{be1}) and (\ref{be2}) in the following section confirms this, see figure \ref{Instability}.
For $\Gamma_2\sim\Gamma_2^{\rm crit}$ the reheating temperature (defined by $\rho_R=\rho_\phi$) is very close to the maximum temperature; both are roughly given by $\tilde{T}_{max}\gg \tilde{T}_R$, and $\rho_R$ rapidly exceeds $\rho_\phi$ by a few orders of magnitude long before $\Gamma=H$. Only a small amount of energy remains in the $\phi$ oscillations until $\Gamma=H$, which roughly occurs at $T=\tilde{T}_R$.
This implies that in this case $\tilde{T}_{max}$ provides the best analytic estimate for both, the maximal and the reheating temperature, which is much larger than $T_R$ defined in (\ref{TR}).
It is clear from (\ref{Gamma2crit}) that this can only be realised if either $\phi$ is rather light or has sizable interactions. It would be extremely interesting to see if there exist realistic models that exhibit this behaviour.

Finally we would like to make a comment on the high temperature behaviour of $\Gamma$. 
 If (\ref{GammaExansion}) contains powers with $n>2$ in the high temperature regime, then it is clear that for sufficiently large $T$, $\Gamma$ always exceeds $H\propto T^2/m_P$ in a radiation dominated universe, possibly leading to a maximal temperature even larger than $\tilde{T}_{max}$. 
  This raises the question whether there are realistic theories in which $\Gamma$ exhibits such behaviour.
Indeed, it has been suggested in \cite{Yokoyama:2006wt} that strong damping due to a term $\Gamma\propto T^4/(m_P^2 m_\phi)$ might solve the cosmic moduli problem. Based on the dimensional arguments given after (\ref{GammaApprox}) we suspect that $\Gamma$ in a renormalisable theory cannot grow with powers $n>2$ in the limit $T\rightarrow \infty$. 
For momenta $\p\sim T$ a scaling $\Gamma\propto T^2/m_\phi$ is rather generic in the regime where $T$ is larger than all masses, but for the zero mode it might even be slower because of the smaller phase space, see (\ref{GammaApproxModel}). For the model suggested in \cite{Yokoyama:2006wt} this was explicitly shown in \cite{Bodeker:2006ij}.
In contrast to that, $\Gamma$ can be a complicated function of $T$ in intermediate temperature regimes, especially near thermal thresholds. In such regimes the procedure of piece-wise approximation by a polynomial introduced in section \ref{Sec:GeneralSol} can of course involve arbitrary positive and negative powers $n$.

\section{An illustrative model}\label{Sec:Model}
Transport in a dense plasma in general is a complicated phenomenon. In the weak coupling regime $\lambda \ll 1$ it can often be understood by modelling the medium as a gas of weakly interacting quasiparticles. 
The properties of these quasiparticles can significantly differ from those of particles in vacuum. Some of them can physically be identified with screened single particle states, and their properties coincide with those of the particles in vacuum in the limit $T\rightarrow 0$. Others have no equivalent at $T=0$ and should be interpreted as collective excitations of the plasma. In general the quasiparticle dispersion relation have a complicated momentum dependence. 
In practice it is common to approximate them by simply replacing the vacuum masses by temperature dependent, but momentum independent "thermal masses", which are obtained by evaluating the dispersion relation at $|\p|=T$.
In a fully thermalised system this often provides a good approximation, as most particles in equilibrium have momenta of order $\sim T$. However, the dissipation of the zero-mode involves $\phi$-quanta with momenta $\ll T$. 
For instance, the decay of a $\phi$-quantum at rest produces daughter particles with momenta $<m_\phi$. In the regime $T\gg m_\phi$ these are infrared or "soft" from the plasma's viewpoint, and $\Gamma$ is highly sensitive to the behaviour of dispersion relations in the plasma in the infrared. 
Hence, the use of such thermal masses is in general not justified during reheating.
This point has been discussed in detail for scalars and fermions with gauge interactions in \cite{Drewes:2013iaa}. 
Here we for simplicity assume that $\phi$ interacts with the plasma of SM particles only via a scalar mediator field $\chi$ with $m_\chi \ll m_\phi$, 
\begin{eqnarray}\label{L}
\mathcal{L}&=& 
\frac{1}{2}\partial_{\mu}\phi\partial^{\mu}\phi
-\frac{1}{2}m_\phi^{2}\phi^{2}
+\frac{1}{2}\partial_{\mu}\chi\partial^{\mu}\chi
-\frac{1}{2}m_\chi^{2}\chi^{2}\nonumber\\
&&-\frac{\alpha}{4!}\phi\chi^3
-\upalpha'm_\phi\phi\chi^2
-\frac{\lambda}{4!}\chi^4
+\mathcal{L}_{\rm bath},
\end{eqnarray}
where $\alpha$ and $\upalpha'$ are dimensionless coupling constants.
$\mathcal{L}_{\rm bath}$ represents the Lagrangian for all other degrees of freedom to which $\chi$ couples directly or indirectly, including the SM fields.
If the $\chi$-dispersion relations are dominated by the quartic self-interaction (rather than terms in $\mathcal{L}_{\rm bath}$), then they are to leading order in $\lambda$ momentum independent, thus we circumvent the difficulty of infrared sensitivity that would occur for general interactions.
Then the $\chi$-dispersion relations for all momenta can be approximated by replacing the vacuum mass $m_\chi$ with the thermal mass $M_\chi$ in all calculations, which is given by\footnote{Another special feature of the quartic self-interaction is that the thermal mass is or order $\propto\sqrt{\lambda}T$. If $\lambda$ were a gauge or Yukawa coupling constant, the thermal mass would scale as $\propto \lambda T$.}
\begin{equation}
M_\chi^2\simeq m_\chi^2 + \frac{\lambda}{24}T^2.
\end{equation}
In principle we should also introduce a thermal mass or "plasma frequency" $M_\phi$ for $\phi$ by evaluating its dispersion relation at zero momentum. Due to the smallness of $\alpha$ we for all practical purposes can take $M_\phi=m_\phi$.\footnote{The analysis in the appendix of \cite{Drewes:2013bfa} suggests that the $\upalpha'm_\phi \phi\chi^2$-interaction generally does not lead to considerable thermal masses, though the behaviour in the infrared is not entirely understood. }

In the limit of vanishing $\chi$-quasiparticle width and neglecting possible collective scalar \emph{luon}-excitations \cite{Drewes:2013bfa}, $\Gamma$ can be analytically approximated as \cite{Drewes:2013iaa}
\begin{eqnarray}
\Gamma &\simeq& \theta(T_2-T)\left(\frac{\alpha^2 m_\phi}{3072\pi^3} + \frac{\alpha^2 T^2}{768\pi m_\phi}\right)
+ \theta(T-T_2)\frac{\alpha^2 m_\phi}{6(2\pi)^4}\frac{T^2}{M_\chi^2}\left(1+{\rm log}\left(\frac{81}{8}\frac{M_\chi}{m_\phi}\right)\right)\nonumber \\
&+&\frac{(\upalpha')^2 m_\phi}{16\pi}\left[1-\left(\frac{2M_\chi}{m_\phi}\right)^2\right]^{1/2}\left(1+2f_B(m_\phi/2)\right)\theta(m_\phi-2M_\chi),\label{GammaApproxModel}
\end{eqnarray}
where $f_B$ is the Bose-Einstein distribution. 
The second line of (\ref{GammaApproxModel}) is the contribution from the $\phi\chi^2$-interaction. 
It is affected by finite temperature effects in two ways. First, induced transitions (``Bose enhancement'') lead to 
the additional term $f_B(m_\phi/2)$. Second, the thermal mass $M_\chi$ modifies the two particle phase space in the square root. 
For $T\ll m_\phi$ the $f_B$-term is negligible and $M_\chi\rightarrow m_\chi$, i.e. the vacuum rate for the decay $\phi\rightarrow\chi\chi$ is recovered. For $T>m_\phi$ the $f_B$-term dominates and can be approximated by $f_B(m_\phi/2)\simeq 2T/m_\phi$, leading to a linear increase of $\Gamma$ with $T$ until reaching a maximum at $T\simeq(\frac{3}{\lambda}(m_\phi^2-4m_\chi^2))^{1/2}$. In this regime one can use (\ref{GammaExansion}) to approximate  $\Gamma\simeq \Gamma_0+\Gamma_1 T/m_\phi$.
For higher temperatures the Bose enhancement looses the competition with the shrinking phase space, 
which suppresses $\Gamma$. The decay eventually is kinematically forbidden when $2M_\chi>m_\phi$.
The appearance of a sharp $\theta$-function that switches the contribution from the $\phi\chi^2$-interaction off for $T>T_c\simeq((m_\phi/2)^2-m_\chi^2)^{1/2}(\lambda/24)^{1/2}$ is a result of the zero width quasiparticle approximation. It gets smeared out once this approximation is dropped and the finite width of $\chi$-quasiparticles as well as contributions to (\ref{PiR}) from
other higher loop diagrams (e.g. vertex and ladder diagrams) are taken into account. Cuts through these can be interpreted as 
contributions to $\Gamma$ from scatterings in which $\phi$-quanta are annihilated, see \cite{Drewes:2013iaa} for a detailed discussion. 
Hence, the contribution from the $\phi\chi^2$-interaction is non-zero even for $2M_\chi>m_\phi$ \cite{Yokoyama:2005dv,BasteroGil:2010pb,Drewes:2013iaa}, 
but is suppressed for $T>T_c$. 
Since in this regime the first line of (\ref{GammaApproxModel}) is unsuppressed, the contribution from the second line is negligible, and using the $\theta$-function is a good approximation for our purpose.
The first line is the contribution from the $\phi\chi^3$-interaction.
It includes contributions from two kinds of processes, decays $\phi\rightarrow\chi\chi\chi$ (first bracket) and scatterings $\phi\chi\rightarrow\chi\chi$ (second bracket), as well as the inverse processes, of course. 
At $T=0$ only the decay processes contribute (first term in the first bracket), but with increasing $T$ they get rapidly overtaken by the scatterings (second term in the first bracket), which are mediated by diagrams of the same order $\mathcal{O}[\alpha^2]$ as the decay. 
In principle the decay also exhibits a kinematic threshold at $3M_\chi>m_\phi$. 
However, at that temperature the scatterings already dominate and we can neglect the temperature dependence of the decay channel, hence the threshold has no significant effect on $\Gamma$ for the parameters we consider \cite{Drewes:2013iaa}.
Moreover, we can neglect the contribution of the vacuum mass $m_\chi$ to $M_\chi$ in this regime.
The scattering contribution grows quadratically for $M_\chi\gtrsim m_\phi$ due to Bose enhancement and the increasing density of scattering partners. Our approximation in this regime is consistent with the result found in the $\phi^4$ model in \cite{Parwani:1991gq}. In the $\phi^4$-model all particles involved in the scattering have the same mass, which is also the case in the model defined by (\ref{L}) in the temperature regime $T\sim T_c$, where $m_\phi\sim M_\chi$. 
At larger temperatures $T\sim T_2$, where $M_\chi\gg m_\phi$, the behaviour differs from that in the $\phi^4$ model. By expanding in $m_\phi/M_\chi$ one can obtain an approximate expression for the damping rate for $T>T_2$, which is given by the second bracket \cite{Drewes:2013iaa}. 
The temperature $T_2\simeq  4.47m_\phi/\sqrt{\lambda}$ is obtained by simply matching the approximate solutions 
given in the first and second bracket, which is sufficient for our purpose.\footnote{The exact expression is $T_2=4M_\phi\sqrt{-W_{-1}[(-16\pi^3)(6561e^2)]/(\pi^3\lambda)}$, where $W$ is the Lambert function.} 
At very high temperatures considerable deviations from (\ref{GammaApproxModel}) may arise because expression has been calculated by inserting resummed propagators in the zero width limit into the leading loop expressions for ${\rm Im}\Pi^R$ in (\ref{PiR}). These do not take into account finite thermal widths, vertex corrections and ``ladder diagrams'', which represent contributions from multiple scatterings that can dominate at large $T$ \cite{Drewes:2013iaa}.

At first we set $\upalpha'=0$. Then (\ref{GammaApprox}) for $T<T_2$ can be parametrised by (\ref{GammaApprox}) with $\Gamma_0=\alpha^2m_\phi/(3072\pi^3)$ and $\Gamma_2=\alpha^2 m_\phi/(768\pi)$. 
This gives
\begin{eqnarray}
T_R&\simeq& 2.5\times 10^{-3}\alpha\sqrt{m_\phi m_P}/g_*^{1/4} \\ 
\tilde{T}_R&\simeq&T_R+3.13\times 10^{-7}\alpha^3\frac{m_P^2}{\sqrt{m_P m_\phi}}g^{-3/4} + \mathcal{O}[\alpha^5] \\
T_{max}&\simeq&3.5\times 10^{-2}\sqrt{\alpha}\left(\frac{m_P m_\phi}{g_*}\right)^{1/4}V_I^{1/8} \\  
\tilde{T}_{max}&\simeq&6.7\times 10^{-3}\alpha\left(\frac{m_P m_\phi}{g_*}\right)^{1/2} \frac{V_I^{1/4}}{m_\phi}
\end{eqnarray}
For the temperature $T_1$ at which finite density corrections start to dominate and $T_2$ where the approximation 
(\ref{GammaApprox}) breaks down we obtain
\begin{equation}
T_1\simeq \frac{m_\phi}{2\pi} \ , \ T_2\simeq 4.47\frac{m_\phi}{\sqrt{\lambda}}. 
\end{equation}
This shows that thermal effects are already crucial for $T<m_\phi$. The condition $T_{max}<T_1$ for the validity of (\ref{Tmax}) can be converted into an upper bound on $V_I$,
\begin{equation}
\frac{V_I^{1/4}}{m_\phi}<\frac{20 \sqrt{g_*}}{\alpha}\sqrt{\frac{m_\phi}{m_P}}.
\end{equation}
For larger $V_I$ we can use (\ref{TtildeMax}) and (\ref{TR}) as long as $\tilde{T}_{max}<T_2$ (i.e. (\ref{GammaApprox}) is valid), which translates into 
\begin{equation}
\frac{V_I^{1/4}}{m_\phi}<\frac{665 \sqrt{g_*}}{\alpha\sqrt{\lambda}}\sqrt{\frac{m_\phi}{m_P}}.
\end{equation}

\begin{figure}
\psfrag{x}{$x$}  
\psfrag{T}{$T/m_\phi$} 
\psfrag{r}{$\rho_R/\rho_\phi$}
\centering
            \includegraphics[width=12cm]{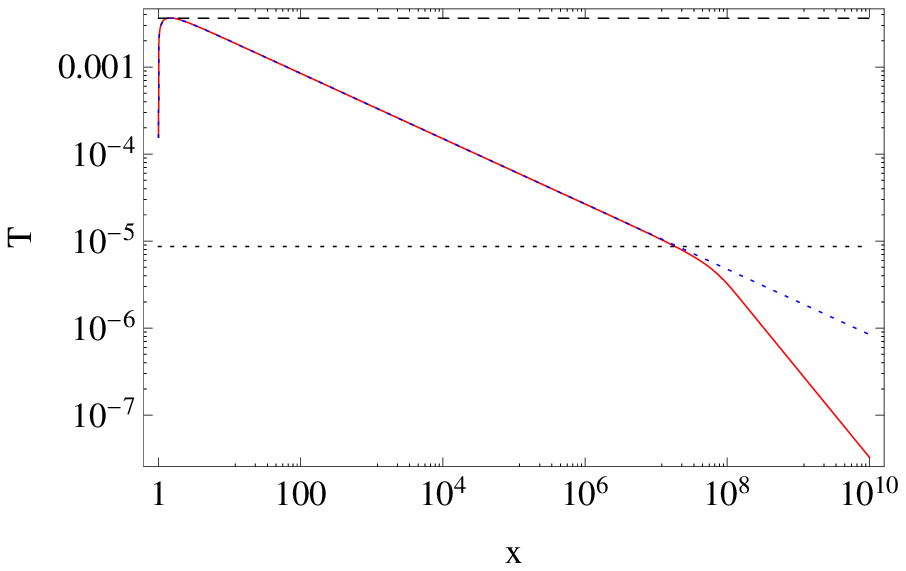}\\
        \includegraphics[width=12cm]{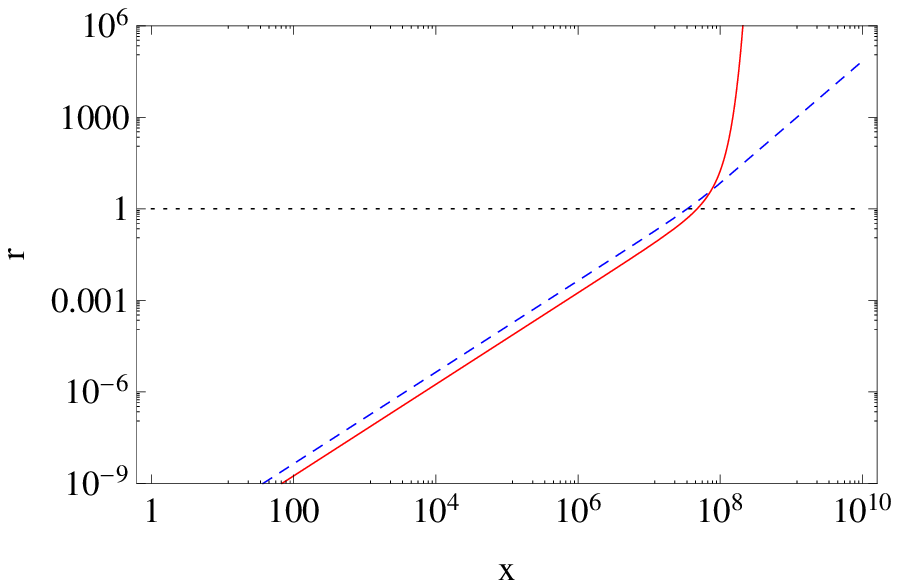}    
        \caption{\emph{Upper panel:}
        The temperature $T$ as a function of $x$ [solid red line] in comparison to the temperature obtained from the approximation (\ref{T1}) [dotted blue line], $T_R$ defined in (\ref{TR}) [dotted black line] and $T_{max}$ defined in (\ref{Tmax}) [dashed black line] for $m_\phi=10^9$ GeV, $m_\chi=10^6$ GeV, $\alpha=10^{-7}$, $\lambda=10^{-3}$, $\upalpha'=0$ and $V_I^{1/4}=10^{10}$ GeV.
        \emph{Lower panel:} The ratios $\rho_R/\rho_\phi$ [red solid line] and $\Gamma/H$ [blue dashed line] as  functions of $x$ for the same parameters.
        \label{NoThermal}}
\end{figure}

\begin{figure}
\psfrag{x}{$x$}  
\psfrag{T}{$T/m_\phi$} 
\psfrag{r}{$\rho_R/\rho_\phi$}
\centering
            \includegraphics[width=12cm]{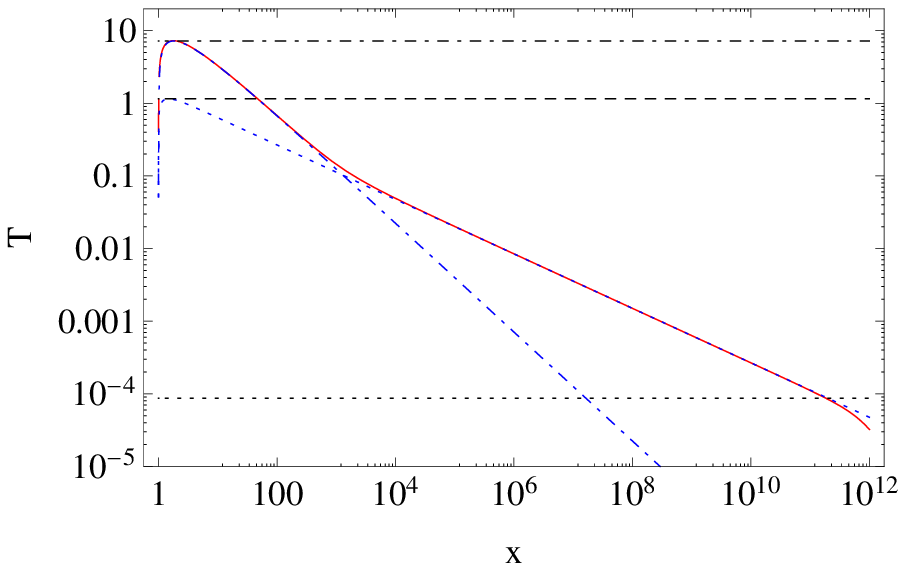}\\
      \includegraphics[width=12cm]{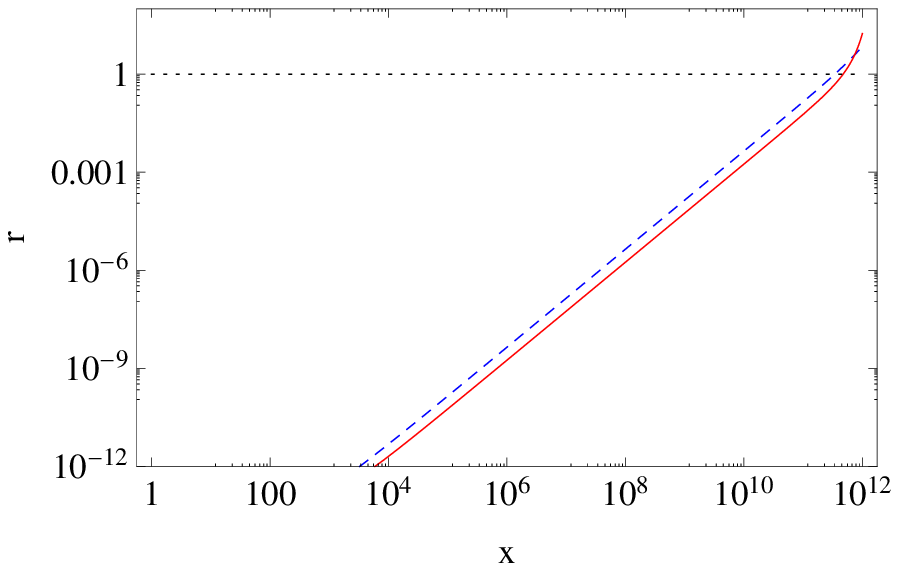}    
        \caption{\emph{Upper panel:}
        The temperature $T$ as a function of $x$ [solid red line] in comparison to the temperature obtained from the approximation (\ref{T1}) [dotted blue line], the temperature obtained from the approximation (\ref{R2}) [dot-dashed blue line], $T_R$ defined in (\ref{TR}) [dotted black line], $T_{max}$ defined in (\ref{Tmax}) [dashed black line] 
        and $\tilde{T}_{max}$ defined in (\ref{TtildeMax}) [dot-dashed black line]         for $m_\phi=10^9$ GeV, $m_\chi=10^6$ GeV, $\alpha=10^{-6}$, $\lambda=10^{-2}$, $\upalpha'=0$ and $V_I^{1/4}=10^{14}$ GeV.
        \emph{Lower panel:} The ratios $\rho_R/\rho_\phi$ [red solid line] and $\Gamma/H$ [blue dashed line] as  functions of $x$ for the same parameters.
        \label{TmaxFig}}
\end{figure}

\begin{figure}
\psfrag{x}{$x$}  
\psfrag{T}{$T/m_\phi$} 
\psfrag{r}{$\rho_R/\rho_\phi$}
\centering
            \includegraphics[width=12cm]{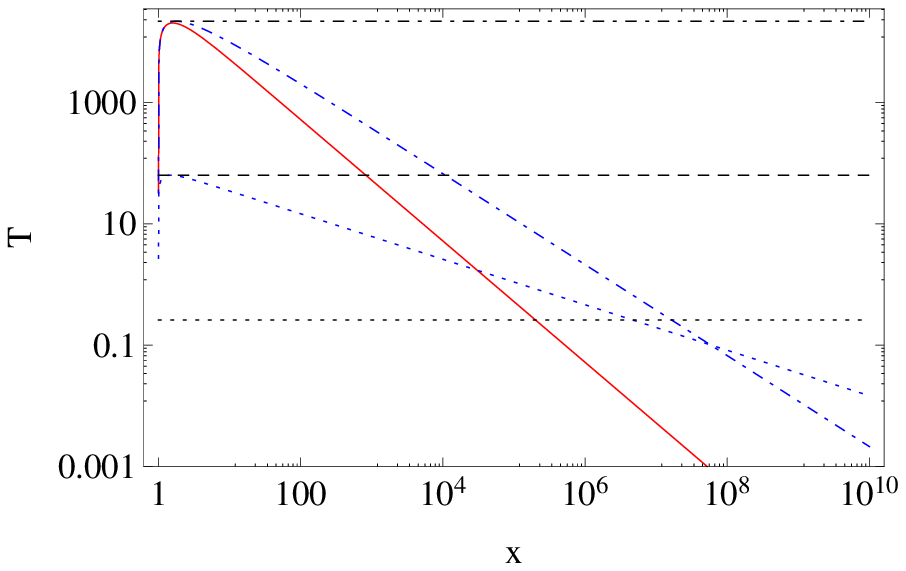}\\
       \includegraphics[width=12cm]{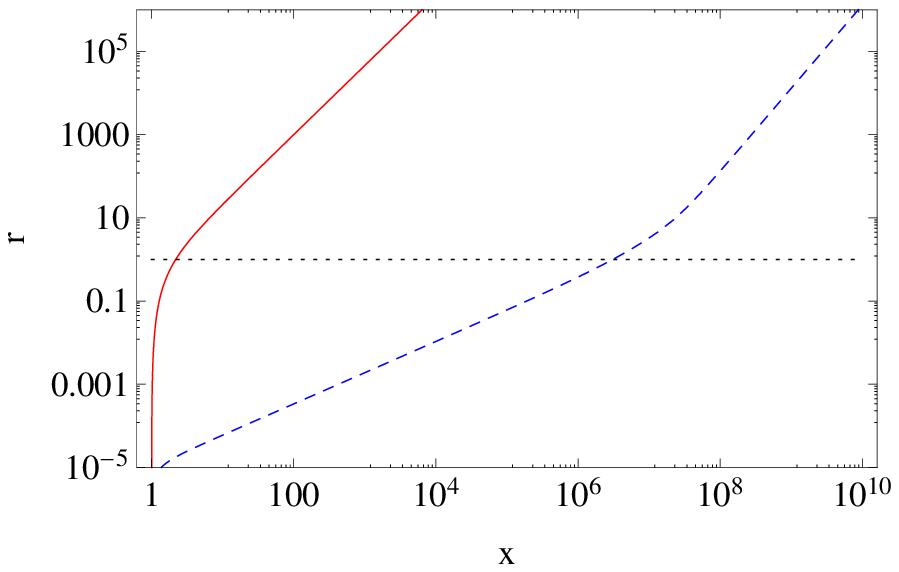}    
        \caption{\emph{Upper panel:}
        The temperature $T$ as a function of $x$ [solid red line] in comparison to the temperature obtained from the approximation (\ref{T1}) [dotted blue line], the temperature obtained from the approximation (\ref{R2}) [dot-dashed blue line], $T_R$ defined in (\ref{TR}) [dotted black line], $T_{max}$ defined in (\ref{Tmax}) [dashed black line] 
        and $\tilde{T}_{max}$ defined in (\ref{TtildeMax}) [dot-dashed black line]     for $m_\phi=10^9$ GeV, $m_\chi=10^6$ GeV, $\alpha= 3\times10^{-3}$, $\lambda=10^{-2}$, $\upalpha'=0$ and $V_I^{1/4}=10^{14}$ GeV.
        \emph{Lower panel:} The ratios $\rho_R/\rho_\phi$ [red solid line] and $\Gamma/H$ [blue dashed line] as  functions of $x$ for the same parameters.
        \label{Instability}}
\end{figure}

\begin{figure}
\psfrag{x}{$x$}  
\psfrag{T}{$T/m_\phi$} 
\psfrag{r}{$\rho_R/\rho_\phi$}
\centering
            \includegraphics[width=12cm]{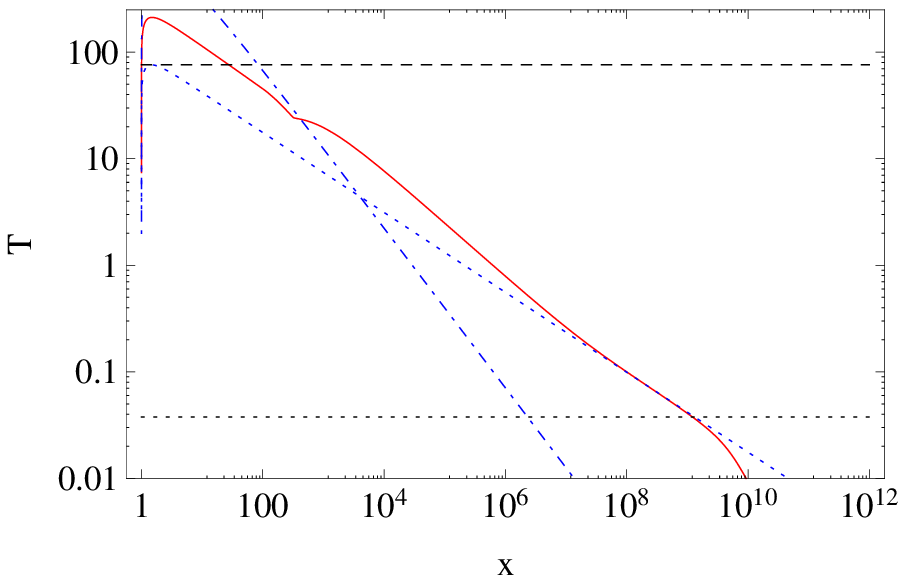}\\
      \includegraphics[width=12cm]{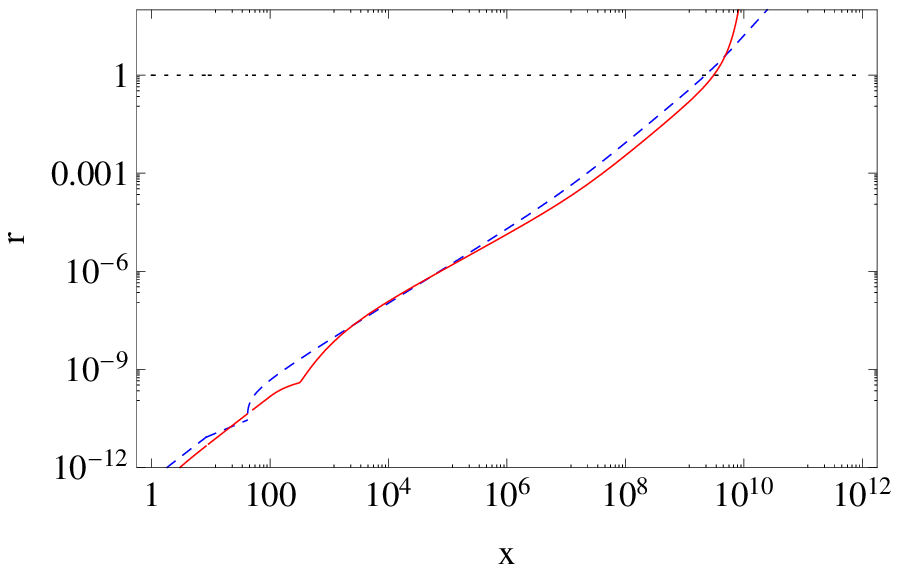}    
        \caption{\emph{Upper panel:}
        The temperature $T$ as a function of $x$ [solid red line] in comparison to the temperature obtained from the approximation (\ref{T1}) [dotted blue line], the temperature obtained from the approximation (\ref{R2}) [dot-dashed blue line], $T_R$ defined in (\ref{TR}) [dotted black line], $T_{max}$ defined in (\ref{Tmax}) [dashed black line] 
        and $\tilde{T}_{max}$ defined in (\ref{TtildeMax}) [dot-dashed black line]     for $m_\phi=10^9$ GeV, $m_\chi=10^6$ GeV, $\alpha=\upalpha' =10^{-5}$, $\lambda=10^{-2}$and $V_I^{1/4}=10^{15}$ GeV.
        \emph{Lower panel:} The ratios $\rho_R/\rho_\phi$ [red solid line] and $\Gamma/H$ [blue dashed line] as  functions of $x$ for the same parameters.
        \label{Thresholds}}
\end{figure}

Figure \ref{NoThermal} shows that for $T_{max}<T_1$ the well-known expressions (\ref{TR}) and (\ref{Tmax}) are good estimates of the reheating temperature and maximal temperature.
The case $T_R<T_1<T_{max}$ is plotted in figure \ref{TmaxFig}. As expected, (\ref{Tmax}) is not valid any more, instead the maximal temperature is given by (\ref{TtildeMax}). At the same time, (\ref{TR}) still holds.
In both cases the radiation dominated era commences at $\Gamma=H$.
Both parameter sets were chosen such that the temperature remains below $T_2$ at all times and (\ref{GammaApprox}) provides a good approximation for $\Gamma$. 
Figure \ref{Instability} shows the case $T_1<T_R<T_{max}$, corresponding to $\Gamma_2>\Gamma_2^{\rm crit}$. 
As suggested by the arguments following (\ref{xcrit}), the radiation dominated era commences much before $\Gamma=H$, and the reheating temperature is very high. To achieve this, we had to choose a rather large coupling $\alpha=3\times10^{-3}$, which may appear unrealistic for the inflaton. However, we have no observational probe of the inflaton potential near its minimum, and it is not obvious that the requirement to have a flat effective potential during the inflationary phase forbids sizable couplings when $\langle\phi\rangle$ is near the minimum. Therefore this possibility cannot be completely ruled out.
In figure \ref{Thresholds} we keep $\Gamma_2\ll \Gamma_2^{\rm crit}$, but choose the initial conditions such that the temperature exceeds $T_2$. In addition, we switch on the other interaction term ($\upalpha'\neq0$).
Then (\ref{GammaApprox}) is no good approximation for $\Gamma$. 
Figure \ref{Thresholds} illustrates two crucial points. First, even for small coupling and $\Gamma_2\ll \Gamma_2^{\rm crit}$ the maximal temperature can be orders of magnitude larger than $m_\phi$ and considerably above $T_{max}$. Second, the non-trivial features in (\ref{GammaApproxModel}), in particular the threshold, leave a visible impact in the temperature evolution. In spite of that, the overall shape of $T$ as a function of $x$ is governed by a power-law once the temperature starts to drop, as expected.

\section{Discussion and conclusions}\label{Conclusions}
We have studied the damping of an oscillating scalar field $\phi$ in the early universe.
Our results can be applied to improve the understanding of the thermal history of the universe during reheating after inflation or the decay of a moduli field.
We included the finite density corrections due to the interactions of $\phi$ with the plasma formed by its own decay products by using a temperature dependent damping rate $\Gamma(T)$.\footnote{For the simplicity of the discussion we make this dependence explicit in this section.} In general (leaving aside specific kinematic regions), $\Gamma(T)$ is larger at high temperature because large occupation numbers enhance the transition probability for bosons and scatterings are more frequent at high density.  
Therefore one can expect that the commonly used expressions for the reheating temperature and maximal temperature in the early universe underestimate the real values of these temperatures.
In general it is difficult to turn this observation into a quantitative statement because $\Gamma(T)$ can be a complicated function of $T$, as different processes (decays and scatterings) contribute to the inclusive rate and the phase space is temperature dependent due to "thermal masses". 
In spite of this, it is possible to understand the time evolution of the temperature analytically by piece-wise approximation of $\Gamma(T)$ by polynomials. 
We used this method to obtain improved analytic estimates of the reheating temperature and maximal temperature in the early universe. It turns out that these can be orders of magnitude larger than the expressions commonly used in the literature. 

This can have a profound effect on the abundance of relics from early epochs of the cosmic history, including Dark Matter. 
In particular, the large plasma temperature allows for thermal production of particles that are heavier than $\phi$ and could not be produced directly from $\phi$-decays for kinematic reasons. 
The abundance of relics that reached equilibrium before freezeout can be determined in the usual way from the freezeout temperature.
The abundance of relics that do not reach thermal equilibrium, on the other hand, is sensitive to the details of the thermal history, which in turn is determined by the functional dependence of $\Gamma(T)$ on $T$.
This dependence can be rather complicated, especially if $\phi$ couples to different fields with multiple interactions, leading to various thermal thresholds. 

To judge whether thermal effects are relevant in a given model, one can calculate the finite temperature damping rate at the temperature $T_{max}$ given by the known expression (\ref{Tmax}), which is commonly used to estimate the maximal temperature. If the vacuum piece of $\Gamma(T_{max})$ is larger than the temperature dependent piece, then (\ref{TR}) and (\ref{Tmax}) are reliable estimates. If the finite temperature piece dominates, then the maximal temperature can be much larger than (\ref{Tmax}). The method of piece-wise approximation of $\Gamma(T)$ by polynomials can be used to derive an improved estimate like (\ref{TtildeMax}). 
The same procedure can be applied to find out whether the standard expression $T_R$ given in (\ref{TR}) provides a good estimate of the reheating temperature at the beginning of the radiation dominated era. If $\Gamma(T_R)$ is dominated by the vacuum piece, then (\ref{TR}) remains valid even if (\ref{Tmax}) does not, otherwise the method of piece-wise approximation can be applied to get a better estimate of the real reheating temperature.
A specific behaviour can arise if the coefficient of the $(T/m_\phi)^2$-term in the expansion of $\Gamma(T)$ in the high temperature limit is larger than $(m_\phi^2/m_P)\times(8\pi^3g_*/90)^{1/2}$. In this case the universe very rapidly enters the radiation dominated phase long before $\Gamma=H$, and the reheating temperature is close to the maximal temperature.

It should, however, be kept in mind that even the improved expressions provided here were derived under a number of simplifying assumptions. In particular, we ignored the highly non-linear phenomenon of non-perturbative particle production that occurs in many models and assumed that the plasma of decay products quickly reaches kinetic equilibrium.

\section*{Acknowledgements}
This work was supported by the Gottfried Wilhelm Leibniz program of the Deutsche Forschungsgemeinschaft and by the National Science Foundation under Grant No. NSF PHY11-25915.
I would like to thank the Kavli Institute for theoretical Physics at the UC Santa Barbara for their hospitality during the work on this project.
\bibliographystyle{JHEP}
\bibliography{all}
\end{document}